\documentclass[conference]{IEEEtran}

\usepackage[colorinlistoftodos,prependcaption,textsize=footnotesize, color=red]{todonotes} 

\usepackage{import}
\usepackage{graphicx}
\usepackage{xcolor}
\usepackage{siunitx}
\usepackage{multirow}
\usepackage{booktabs}
\usepackage{balance}
\usepackage{tikz}
\usepackage{import}
\usepackage{paralist}
\usepackage{subcaption}
\usepackage{booktabs}

\usepackage{xspace}

\usepackage{makeidx}

\begin{document}

\title{Mitigation of Flooding and Slow DDoS Attacks in a Software-Defined Network}

\author{
    \IEEEauthorblockN{Thomas Lukaseder$^1$, Shreya Ghosh$^2$, Frank Kargl$^1$}\\
	\IEEEauthorblockA{$^1$Institute of Distributed Systems\\
	Ulm University, Germany\\
	\{firstname\}.\{lastname\}@uni-ulm.de}\\
	\IEEEauthorblockA{$^2$Department of Electronics and Communication Engineering\\
    Heritage Institute of Technology, India\\
	shreya.ghosh.ece19@heritageit.edu.in}
}

\maketitle

\section{Introduction}

Distributed denial of service (DDoS) attacks are a constant threat for services in the Internet. This year, the record for the largest DDoS attack ever observed was set at 1.7~Tbps. Meanwhile, detection and mitigation mechanisms are still lacking behind. Many mitigation systems require the assistance by the victim\,---\,or the victim's administrator themself has to become active to mitigate attacks. We introduced a system that can detect attacks, identify attackers, and mitigate the attacks purely within the network infrastructure. With the improved flexibility of software-defined networks, new possibilities to mitigate such attacks can be implemented. In addition to our short paper on the mitigation of reflective DDoS attacks on LCN 2018~\cite{drdos}, we also like to demonstrate our work on mitigating flooding attacks presented at LCN 2017~\cite{sdndos} and our mitigation of slow DDoS attacks~\cite{slowdos}. In our demo, we show how these systems can be combined and how they work when faced with such different attacks.

This paper is organized as follows: Section~\ref{sec:description} describes the attacks that will be mitigated in this demo while Section~\ref{sec:arch} contains information about the mitigation system as a concept. Section ~\ref{sec:setup} explains the inner workings of the implementation while Section~\ref{sec:demo} illustrates the procedure of the demo from an observer perspective followed by the requirements in Section~\ref{sec:requirements}.

\section{Attack Description}
\label{sec:description}

There is a large variety of DDoS attacks. For the purpose of this demo, we narrow it down to a variety of flooding attacks and some slow attacks. For one, the \emph{SYN flooding attack}. This attack exploits the connection limit of a server. When TCP connections are initiated by a client with a SYN packet, the server needs to save the state of the connection in a table. As soon as the connection is established this entry can be removed. The table to save these connections is of a fixed size, therefore, only a set amount of connections can be opened at the same time. During normal operation, this is not a problem as the time for connections to be opened is rather short. However, during a SYN flooding attack, vast amounts of connections are opened by sending SYN packets with spoofed source IP addresses. By not sending a SYN-ACK packet, these connections are never fully established and remain in the table until their regular time out, effectively blocking the server from benign connection attempts.

In a \emph{TLS flooding attack}, the attacker exploits the fact that in older TLS and SSL versions, the handshake\,---\,based on the Diffie-Hellmann key exchange protocol\,---\,is quite resource intensive on the server with a much lower load on the client. Therefore, mass renegotiations of TLS sessions overloads the server.

\emph{HTTP flooding attacks} are a classical example of a brute force attack. The attacker looks for a costly operation at the victim that can be triggered by a simple HTTP request, e.g. an operation requiring database access, search operation, or running a pathfinding algorithm. We simulate this on our demo site with the calculation of a bcrypt hash. When such a resource intensive request possibility is found, the attackers send as many requests as possible.

The \emph{slow attacks} are working very differently. Instead of sending requests as fast as possible, they send as slow as possible. The \emph{slowloris} or \emph{slow header attack} sends HTTP get requests in as many packets as possible (by splitting the header in several packets) and then sending these at a very low packet rate (e.g. 0.5 Hz; 30s between packets). The attacker needs to keep as many concurrent connections open as the server provides. As the attacker only needs to send a few packets per minute per connection, only little resources are necessary to render the server inaccessible. The \emph{slow body attack} or \emph{slow POST attack} works in similar ways. However, instead of sending the header as slow as possible the body of a POST message is split into as many packets as possible.

Slow attacks are hard to detect in the network without access to the victim machine as the attackers are behaving according to specification and could easily be mistaken for clients with bad network connections.

\section{Defense Architecture}
\label{sec:arch}

To defend against DDoS attacks, several steps have to be undertaken. For one, the attack has to be observed. Then, the attack has to be classified to find the matching defense mechanism. In a third step, the individual attackers need to be identified and in a last step, the attackers need to be excluded from the network (blocked) to mitigate the attack.

The first step can be done by the victim reporting the attack. However, in many networks the attack victim and the service provider running the mitigation system do not necessarily cooperate. Furthermore, an attack victim that is unreachable due to an attack might not be able to report the attack anymore. Therefore, we observe the attack target from the outside by measuring the round trip time (RTT) of regular requests to the victim. If the RTT consistently exceeds a threshold typical for that service, the attack classification is launched.

In the attack classification phase, several outcomes are possible. For one, the rise in RTT can be due to unrelated issues on the server and no attack can be found or the attack is not known to the system. However, if the traffic can be attributed to typical attack patterns (e.g. high amounts of SYN packets combined with a traffic spike (SYN flooding) or many evenly spaced packets sent with a very low packet rate (slowloris), the attack can be found and classified.

In the identification phase, attackers need to be identified dependent on the classified attack. This is done by observing the typical behavior of attackers.

In the last step\,---\,the mitigation phase\,---\,the list of now identified attackers is sent to the SDN controller that in turn sets rules in the switch to block these clients.

It is imperative to keep on running all steps in parallel and to update the later phases with newfound insights. The first step needs to run to recognize the successful mitigation of the attack, the second step needs to run to recognize if a second attack type might be running and needs to be mitigated and the third step needs to keep on running to see if new attackers can be found in the network. More details on the mitigation architecture can be found in the original publications.

\section{Setup}
\label{sec:setup}

The original setup running in real hardware in our research network was ported to a portable implementation based on virtual machines and virtual network equipment.

\begin{figure}
    \includegraphics[width=\columnwidth]{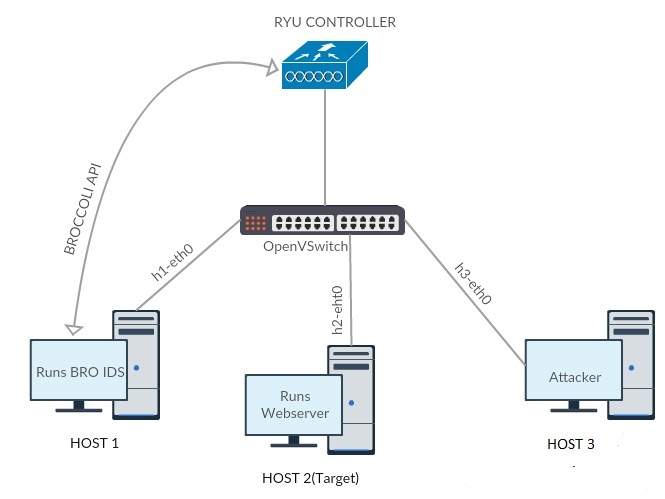}
    \caption{Setup in Mininet.}
    \label{fig:setup}
\end{figure}

The emulations setup can be seen in Figure~\ref{fig:setup} and has been conducted using a Linux Virtual Machine packaged with Mininet 2.2.2. It is configured with a Host-only adapter and a NAT Adapter, and has two network interfaces. The RYU Controller is used as SDN controller. It is connected to an OpenVSwitch (SDN Switch) version 2.5.0. The protocol used is OpenFlow1.3.  To emulate the real attacker scenario, three hosts are connected to the OpenVSwitch. Each host is provided with Internet connectivity. They are all connected by 13 Gbit/s links.

Host 1 runs the Bro Network Security Monitor in version 2.5.4, that is configured to observe all traffic on the network interface and detect flooding or slow attacks according to our previous publications~\cite{sdndos,slowdos}, and is connected to the OpenVSwitch. Bro communicates its events to the RYU controller via the Bro Communication Library (BROCCOLI) API.

Host 2 is the target host that runs the Apache web server and the application. The running application calculates the response time of the server when it receives GET requests. 

Host 3 is used to emulate an attacker and attacks target machine Host 2. All packets that are transferred from the attacker host to the target host is mirrored to Host 1 running Bro. This enables Bro to observe all the packets inside the network and differentiate between benign traffic and harmful traffic.

The attacking tools in use can be seen in Table~\ref{tab:attacks}. We use slowloris, slowHTTPtest and hping3. Where necessary, these tools were adapted to enable spoofing of different source IP addresses. This way, several hundred attack clients can be simulated per attack.

\begin{table}
    \centering
    \caption{Attacks and tools used in the demo.}
    \label{tab:attacks}
    \begin{tabular}{lll}
    \toprule
    \textbf{Attack group}             & \textbf{Attack}  & \textbf{Tool}           \\
    \midrule
    \multirow{2}{*}{Slow Attacks}     & Slow Header      & slowloris               \\
                                      & Slow Body        & slowHTTPTest            \\
    \multirow{3}{*}{Flooding Attacks} & HTTP Flooding    & \multirow{3}{*}{hping3} \\
                                      & SYN Flooding     &                         \\
                                      & SSL-TLS Flooding &                      \\  
    \bottomrule
    \end{tabular}
\end{table}

\begin{figure*}
    \centering
    \includegraphics[width=\textwidth]{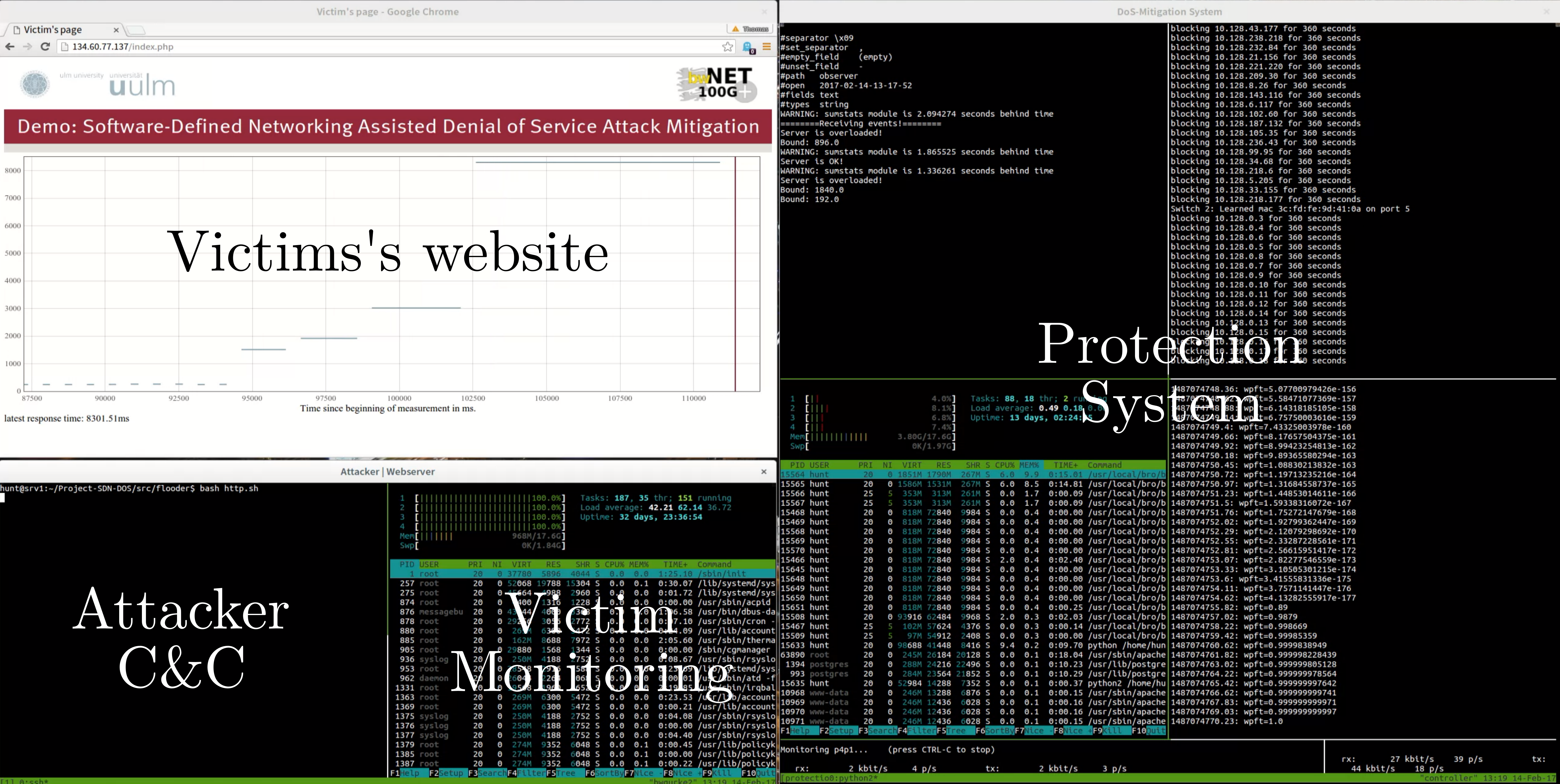}
    \caption{Demo from the observer perspective.}
    \label{fig:demo}
\end{figure*}

\section{Demo}
\label{sec:demo}

Figure~\ref{fig:demo} shows the demo from the observer perspective. In the demo, the performance of the victim server is observable using a website on this very server (upper left corner). Within this website, the response time for the GET requests are being visualized by the length of lines in the diagram. The measurement is done by javascript AJAX requests in the background. The requests go to a second page that calculates a bcrypt hash to show a considerable response time even when the server is not under attack for visualization purposes. The lower left corner shows the terminal window of the attacker command and control system. Here, the attacks will be started. Right next to it is the monitoring of the victim's machine. On the right, the protection system is shown. The different windows of the protections system show (from top left to lower right) the output of Bro (including the message whether an attack was observed or not), the output of Ryu (including the list of blocked IP addresses), the system monitor of the protection system and an output of the internal parameters used by the protection system.

In a first run, the attacks will be run without the protection system active to show their impact on the victim. The different attacks show different results. While a SYN flooding attack or a slowloris attack will result in a sudden unreachability of the server as soon as the maximum amount of concurrent connections is reached, an HTTP flooding attack will lead to a slow rise of the response time. Meanwhile, the system monitor of the victim shows very different impact on the CPU usage. While the SYN flooding and slowloris attack leads to the CPU usage to go down (as legitimate connection attempts are not patched through to the CPU anymore) the HTTP flooding attack shows maximum CPU usage which in turn leads to the unreachability of the victim. The server is flooded with requests from the attacker (Host 3) that leads to considerable stress on the CPU of the server. The server is unable to respond to these requests due to the huge traffic load. As a result, the response time increases which is reflected by the increased lengths of the lines on the diagram. It can also be observed that the different attacks take different times to be effective and the server takes different times to recover.

Before the second run, the protection system is activated. When the protection system is triggered, it starts the mitigation. For any identified attacker, it sends an event to the RYU controller that blocks the attacker from sending any more data packets. The attackers cannot reach the target at this stage. As a result, the server becomes available again and the response time improves. This is also reflected in the diagram as the length of the lines indicating the response time become shorter. Thus, the victim is no longer affected by the attack demonstrating the in-network security provided to the victim.

\section{Requirements for the Demo}
\label{sec:requirements}

\begin{itemize}
    \item A table with space for one laptop and one monitor
    \item Power supply for the laptop and monitor
    \item A monitor with VGA or Mini DisplayPort
    \item Internet connection
    \item Poster board
    \item Setup-time: 1 hour
\end{itemize}

\section*{Acknowledgment}

This work was supported in the bwNET100G+ project
by the Ministry of Science, Research and the Arts Baden-
W\"urttemberg (MWK). The authors alone are responsible for
the content of this paper.

\bibliographystyle{IEEEtranS}
\bibliography{literature}
\end{document}